\documentstyle[11pt,aaspp4,epsfig]{article}

\begin{document}
\newcommand{\bref}[1]{(\ref{#1})}
\newcommand{\bea}[0]{\begin{eqnarray}}
\newcommand{\eea}[0]{\end{eqnarray}}
\newcommand{\beann}[0]{\begin{eqnarray*}}
\newcommand{\eeann}[0]{\end{eqnarray*}}
\newcommand{\sign}[0]{{\rm sign}}
\newcommand{\be}{\begin{equation}}
\newcommand{\ee}{\end{equation}}
\newcommand{\dt}{\Delta t}
\newcommand{\dx}{\Delta x}
\newcommand{\dr}{\Delta r}
\newcommand{\dphi}{\Delta \phi}
\newcommand{\dy}{\Delta y}
\newcommand{\dz}{\Delta z}
\newcommand{\real}{R}
\newcommand{\Mu}{{\rm M}}
\newcommand{\nn}{\nonumber}
\newcommand{\hlf}{\frac{1}{2}}
\newcommand{\frt}{\frac{1}{4}}
\newcommand{\pdr}{\frac{\partial}{\partial r}}
\newcommand{\pdt}{\frac{\partial}{\partial t}}
\newcommand{\pdp}{\frac{\partial}{\partial \phi}}

\title{Rossby Wave Instability of
Thin Accretion Disks -- III. Nonlinear Simulations}

\author{H. Li\altaffilmark{1,2}, S.A. Colgate\altaffilmark{1},
B. Wendroff\altaffilmark{3} and R. Liska\altaffilmark{4}}

\altaffiltext{1}{Theoretical
Astrophysics, T-6, MS B288, Los
Alamos National Laboratory, Los
Alamos, NM 87545. hli@lanl.gov}
\altaffiltext{2}{Applied Physics Division, Los
Alamos National Laboratory, Los Alamos, NM 87545}
\altaffiltext{3}{T-7, Los Alamos National Laboratory, Los Alamos, NM 87545}
\altaffiltext{4}{Faculty of Nuclear Sciences and Physical Engineering,
Czech Technical University, Prague, Czech Republic}

\begin{abstract}

We study the nonlinear evolution of the Rossby wave instability in
thin disks using global 2D hydrodynamic simulations. The detailed 
linear theory of this nonaxisymmetric instability was developed
earlier by Lovelace et al. and Li et al., who found that the
instability can be excited when there is an extremum in the radial 
profile of an entropy-modified version of potential vorticity. 
The key questions we are addressing in this paper are: 
(1) What happens when the instability becomes nonlinear? Specifically, 
does it lead to vortex formation? (2) What is the detailed behavior of
a vortex? (3) Can the instability sustain itself and can the vortex
last a long time? Among various initial equilibria that we have
examined, we generally find that there are three stages of the disk 
evolution: (1) The exponential growth of the initial small amplitude
perturbations. This is in excellent agreement with the linear theory;
(2) The production of large scale vortices and their interactions
with the background flow, including shocks. Significant accretion is 
observed due to these vortices. (3) The coupling of Rossby
waves/vortices with global spiral waves, which facilitates further 
accretion throughout the whole disk. Even after more than 20
revolutions at the radius of vortices, we find that the disk maintains 
a state that is populated with vortices, shocks, spiral waves/shocks, 
all of which transport angular momentum outwards.
We elucidate the physics at each stage and show that there is an
efficient outward angular momentum transport in stages (2) and (3)
over most parts of the disk, with an equivalent Shakura-Sunyaev
angular momentum transport parameter $\alpha$ in the range
from $10^{-4}$ to  $10^{-2}$.  By carefully analyzing the flow
structure around a vortex, we show why such vortices prove to be
almost ideal ``units'' in transporting angular momentum outwards,
namely by positively correlating the radial and azimuthal velocity components.
In converting the gravitational energy to the internal energy, we find
some special cases in which entropy can remain the same while angular 
momentum is transported. This is different from the classical $\alpha$
disk model which results in the maximum dissipation (or entropy production).  
The dependence of the transport efficiency on various physical parameters are
examined and effects of radiative cooling are briefly discussed as well.
We conclude that Rossby wave/vortex instability is an efficient,
purely hydrodynamic mechanism for angular momentum transport in thin disks,
and may find important applications in many astrophysical systems.

\end{abstract}

\section{Introduction}

Understanding the physics of accretion disks has 
remained a great challenge
in astrophysics for decades. Matter has to lose angular momentum
in order to fall deeper into a gravitational potential well.
The release of the gravitational binding energy then becomes one
of the most powerful energy sources in the universe. 
Various models for angular momentum transport have been proposed,
including those having a purely radial transport (i.e., within the disk)
and those using outflows (e.g., MHD jets).  
One promising mechanism for removing angular momentum locally within
the disk is via MHD turbulence
in disks (\cite{bh98}). Disks must be made of relatively hot and
sufficiently ionized plasma for this mechanism to operate, however,
because a strong coupling between magnetic fields and plasma is required. 
On the other hand, there are several types of astrophysical disks where such 
conditions are not fully satisfied. Thus, a purely hydrodynamic means of
angular momentum transport is still needed (see \cite{paplin95} for a
recent review).    

In two previous papers, Lovelace et al. (2000, hereafter Paper I)
and Li et al. (2000, hereafter Paper II) have presented a detailed
analysis of the linear theory
of a global, nonaxisymmetric hydrodynamic instability in thin (2D) disks. 
The disk becomes unstable when the conditions of Rayleigh's
inflection point theorem are violated, which is indicated by the 
radial profile of a
key function 
${\cal L}(r) \equiv (\Sigma\Omega/\kappa^2) S^{2/\Gamma}$. 
This function is an
entropy-modified version of potential vorticity. Here, $\Sigma(r)$ is
the surface mass density of the disk, $\Omega(r)$ the angular rotation
rate, $S(r)$ the specific entropy, $\Gamma$ the adiabatic index, and
$\kappa(r)$ the radial epicyclic frequency.
It has been shown that a sufficient variation of pressure over a length
scale that is a few times the disk thickness can cause the disk to be
unstable to nonaxisymmetric perturbations even though the disk is still
locally {\em stable} to axisymmetric perturbations according to the
so-called Rayleigh determinant or the Solberg-Hoiland criterion when
the pressure effects are included (Paper II).     
The linear theory shows that the unstable modes have a dispersion
relation similar to that of Rossby waves in atmospheric studies (Paper
I). The term ``Rossby wave instability'' (RWI) was introduced in
that paper. The dependence of RWI on various physical
parameters has been examined in Paper II and its relations with other
hydrodynamic instabilities, especially the Papaloizou \& Pringle
instability (\cite{pp85}), have been discussed in Paper II as well.
More generally, since the pioneering work by Papaloizou \& Pringle
(1985), nonaxisymmetric instabilities in disks have received an
enormous amount of attention with various degrees of success. 
In particular, the important role of ``vortensity'' (vorticity divided
by surface density) in determining the stability of the disk was
first discussed in Lovelace \& Holfeld (1978) and was studied in
greater detail in Papaloizou \& Lin (1989). 
 Nonaxisymmetric
convective (in the vertical direction) instability is also explored
in Lin, Papaloizou, \& Kley (1993). Sellwood \& Kahn (1991)
demonstrated that a ``narrow groove'' in the angular momentum density
profile can be very unstable. Interestingly, Toomre (1981)
had already noticed certain unstable modes associated with
disk edges (termed ``edge modes'').  Even though these earlier studies
have mostly used a homentropic equation of state (i.e., effects due to 
an entropy gradient were usually not present), they are, in general,
consistent with the above mentioned criterion by having
an inflection point in the radial profile of the key function
${\cal L}(r)$.

There have been several other recent studies of the role of
Rossby waves in accretion disks. Sheehan et al. (1999)
performed a linear analysis of the generation of Rossby waves and
their propagation in protoplanetary disks. They furthermore
speculated on the possibility of forming vortices and zonal jets
(similar to planetary atmosphere dynamics). Using extensive 
nonlinear disk simulations, 
Godon \& Livio (2000) have investigated the formation of
vortices in a viscous, compressible flow and 
Klahr \& Bodenheimer (2000) have shown that vortices can be 
produced by a global radial entropy gradient, possibly via a baroclinic 
instability from which angular momentum is transported outwards 
with an efficiency at $\alpha \sim 10^{-3}$ level. 

Besides the possible role of Rossby waves/vortices, another
important purely hydrodynamic angular momentum transport mechanism
is through spiral waves/shocks. This has been proposed for
systems such as cataclysmic variables (CVs; 
see \cite{spruit91} for a review)
and accreting neutron stars (\cite{m84}), where
the nonaxisymmetry in the disk flow is caused by external torques
acting on the disk either from a close companion or an asymmetric
central rotating body (such as the magnetosphere of a neutron star). 
In the linear theory analysis of
Tagger \& Pellat (1999), a possible
mechanism of coupling Rossby waves with spiral (density) waves through
the corotation resonance was briefly discussed.

In this paper, we use extensive 2D hydrodynamic simulations to study 
the nonlinear evolution of RWI, based on the knowledge we have
gained from the linear theory analysis. 
After a brief description of our 2D hydro code, 
we show how the initial states of disk simulations are set up 
in \S \ref{sec:lwlf},
  We then present the simulation results 
in \S \ref{sec:result1} and \S \ref{sec:result2}.
In \S \ref{sec:discuss}, we discuss several important physical
issues that are associated with this work. Conclusions are given
in \S \ref{sec:conclu}.

\section{The 2D Hydrodynamic Model}
\label{sec:lwlf}

Several simplifying assumptions are employed in this study. The disk
is assumed to be geometrically thin so that the
hydro equations can be reduced to 2D with vertically integrated
quantities. The effects due to magnetic field and self-gravity are
omitted and the Newtonian potential is used throughout our
simulations. The disk is
treated as an isolated system so that there is no externally supplied
mass inflow, but mass outflow through the disk
radial boundaries (both sides) is allowed.

\subsection{The differential equations}

The Euler equations are the governing equations for our
2D (in ${r,\phi}$ plane), inviscid
disk flow with a central gravitating object. 
The usual variables in Euler equations are
$\underline{u} = \{\Sigma, v_r, v_\phi, E \}$, where $v_r, v_\phi$ are
the radial and azimuthal velocities, respectively, $E$ is the total
energy $E = P/(\Gamma-1) + 0.5\times \Sigma (v_r^2 + v_\phi^2)$, 
$\Gamma$ is the ideal gas adiabatic index, and $P$
is the vertically integrated pressure. 
Here, we choose to use a new set of variables
$\underline{u} = \{r\Sigma, r\Sigma v_r, r^2 \Sigma v_\phi, r E \}$.
One advantage of this choice is to eliminate the
nonzero source term in the angular momentum equation. 
Consequently, the Euler equations in cylindrical coordinates become
\begin{equation}
\label{eq:euler}
\frac{\partial \underline{u}}{\partial t} 
+ \frac{\partial f(r,\underline{u})}{\partial r} 
+ \frac{\partial g(r,\underline{u})}{\partial \phi} 
+ S(r,\underline{u}) = 0
\end{equation}
where
\begin{equation}
\underline{u} = 
\left( \begin{array}{c} r \Sigma \\ r \Sigma v_r\\ r^2\Sigma v_\phi \\ r E
 \end{array}\right) ~~~~,
\end{equation}
and 
\begin{equation}
f = 
\left( \begin{array}{c} r \Sigma v_r \\ 
r \Sigma v_r^2 \\ r^2 \Sigma v_r v_\phi \\ (r E + r P) v_r
 \end{array}\right) ~~~~,
\end{equation}

\begin{equation}
g = 
\left( \begin{array}{c} \Sigma v_\phi \\ \Sigma v_r v_\phi \\ 
r \Sigma v_\phi^2 + r P \\ (E + P) v_\phi
 \end{array}\right) ~~~~,
\end{equation}
\begin{equation}
S = 
\left( \begin{array}{c} 0 \\ r dP/dr - \Sigma(v_\phi^2 - 1/r)
 \\ 0 \\ \Sigma v_r/r
\end{array}\right) ~~~~.
\end{equation}
\noindent The $\Sigma/r$ term in $S$
is the normalized central 
gravitational potential. 
The zero components of
$S$ express the conservation of mass and angular momentum.

\subsection{The Initial Conditions}
\label{sec:setup}

We study the evolution of RWI by first setting up the initial
equilibria which are steady, axisymmetric
($\partial/\partial \phi = 0$) and with zero
radial velocity ($v_r = 0$). Following the analysis in Paper II,
we first specify the surface density $\Sigma(r)$
and temperature $T(r)$ if the disk does not have a constant
entropy initially. 
As the main focus of this paper is to study the
nonlinear evolution of the linear instability found in Paper I and II,
we follow the initial setups in Paper II and concentrate on two types
of initial equilibrium. One has a Gaussian-shaped ``bump'' in pressure
 and the other has a step-jump in
pressure. For the sake of completeness, we recap the
functions used to describe the bump/jump (Paper II),
\begin{equation}
\label{eq:gau-rho}
Bg(r) = 1 + ({\cal A}-1)
\exp\left[-{1\over
2}\left(\frac{r-r_0} {\Delta
r}\right)^2\right]  ~,
\end{equation}
for the Gaussian bump and 
\begin{equation}
\label{eq:sj-rho}
Bj(r) = 1 + \frac{\cal A}{2}
\left[\tanh\left(\frac{r-r_0}
{\Delta r}\right) + 1 \right]~~,
\end{equation}
for the step jump. Quantities ${\cal A}$ and $\Delta r$  measure
the amplitude and width of the bump/jump respectively, 
and $r_0$ is radial location of the bump/jump. 
Specifically, we have considered 4 basic types,
\begin{equation}
{\rm HGB:} \left\{
\begin{array}{lcl}
\rho(r) & = & \rho_0~ Bg \nonumber \\
P(r) &  = & P_0 \left[ \rho(r)/\rho_0 \right]^{\Gamma}
\end{array}
\right.~~,
\end{equation}

\begin{equation}
{\rm NGB:} \left\{
\begin{array}{lcl}
\rho(r) & = & \rho_0~(r/r_0)^{-3/4}  \nonumber \\
T(r) & =& T_0~(r/r_0)^{-3/4}~Bg  \nonumber \\
P(r) &  = & \rho_0 T_0~(r/r_0)^{-3/2}~Bg
\end{array}~~,
\right.
\end{equation}

\begin{equation}
{\rm HSJ:} \left\{
\begin{array}{lcl}
\rho(r) & = & \rho_0~Bj  \nonumber \\
P(r) &  = & P_0~\left[ \rho(r)/\rho_0 \right]^{\Gamma}
\end{array}
\right.~~,
\end{equation}

\begin{equation}
{\rm NSJ:} \left\{
\begin{array}{lcl}
\rho(r) & = & \rho_0~(r/r_0)^{-3/4}~Bj \nonumber \\
T(r) & =& T_0~(r/r_0)^{-3/4}~Bj  \nonumber \\
P(r) &  = & \rho_0 T_0~(r/r_0)^{-3/2}~Bj^2
\end{array}
\right.~~,
\end{equation}
which are named the homentropic Gaussian bump (HGB), nonhomentropic
Gaussian bump (NGB), homentropic step jump (HSJ), and nonhomentropic
step jump (NSJ) cases, respectively. 
These equations have an overall normalization so 
that $c_{s0}^2 = \Gamma P_0/\Sigma_0 = \Gamma T_0 = 0.01 v_{\phi k}^2(r_0)$.
Note that even though we have used either $P \propto \Sigma T$ or
$P \propto \Sigma^\Gamma$ in obtaining the
initial pressure distributions,
we only use $P \propto \Sigma T$ as the equation of state
during the subsequent disk evolution.
For a given $\Sigma(r)$ and $P(r)$, we use the radial
force balance to calculate the azimuthal velocity $v_{\phi}$, which is
very close to Keplerian velocities except the slight modification by
the pressure gradient (Paper II). 

To be consistent with the 2D approximation, we require
that the length scale of the pressure variation 
$L_p = \Gamma P/|dP/dr|$ (in units of $r_0$) is larger than the disk
scale height $\sim c_s/v_{\phi}$ at $r_0$. 
Note that $L_p$ is always larger than $\Delta r$.
For example, using equation 
(\ref{eq:sj-rho}), if $\Delta r/r_0 = 0.05$ and
${\cal A} = 0.65$, the minimum of $L_p/r_0$ is $\sim 0.2$, which
is twice the thickness of disk when $c_s/v_\phi$ at $r_0$ is $0.1$. 
In the following runs, we have used both $\Delta r/r_0 = 0.05$ and $0.1$.
The amplitude ${\cal A}$ is the main parameter to be varied.
As discussed in Paper II, the RWI growth rate $\gamma$ of RWI is a strong
function of $c_s$ and $L_p$, and the growth rate 
$\gamma$ becomes a fraction ($\sim 0.2$) of the Keplerian rotation 
$\Omega(r_0)$ when $L_p$ is 
$\sim 2\times$ the disk thickness. This is because the pressure
gradient is the only force available to perturb the Keplerian flow in
an inviscid disk.

There are several considerations that go into constructing these
different types. The disks of HGB and HSJ types start with a single
entropy for the whole disk. They are relatively ``clean'' systems, and 
so help us to determine
whether an entropy gradient is needed in the development of RWI.
The linear theory in Paper II says $dS/dr$ can be zero for RWI to
grow. We can test it and compare them with the cases of 
NGB and NSJ where a background radial gradient in entropy is present.
The power law dependence of $\Sigma(r)$ and $T(r)$ in NGB and NSJ 
mimics the distribution from a steady state $\alpha-$disk model,
although the code can handle an arbitrary slope. Furthermore,
the bump in NGB is in temperature only whereas NSJ has a jump
in both density and temperature.

\subsection{Description of the numerical method}
\label{sec:num}

The system (\ref{eq:euler}) is integrated using a
dimensional-splitting method. Furthermore, a local
co-moving frame for the $\phi$ sweeps is employed. 
It was observed by Masset (2000) that this reduces the computing time
greatly over a fully two-dimensional method, the reason
being that $v_\phi$ is much greater than the sound speed over the
whole disk. Indeed, with this innovation we have been able to make
hundreds of 
computer runs on a workstation testing various configurations.
We have modified the method by Masset in several aspects 
for our codes and the full details of the numerical method 
are given in the appendix.

\subsection{Boundary conditions and Numerical Dissipations}

The boundary condition along the azimuthal direction is periodic.
It proves to be very difficult in determining what is the best
radial boundary condition. Since we are simulating a small part of the
whole disk, ideally, we want to use the propagating sound wave
conditions to minimize the possible reflection effects  at
the boundaries.  Let $r_1$ and $r_2$ be the inner and outer 
disk boundaries, respectively.
When ``signals'' are generated near $r_0$ and propagate with
sound  speed $c_s$ both inwards and outwards, it takes roughly
$(r_0-r_1)/c_s$ and $(r_2 - r_0)/c_s$ to reach the inner and outer
boundaries, respectively. So, if $c_s/v_\phi(r_0) \sim 0.1$, then
after about 2 revolutions at $r_0$, signals would have reached the inner
boundary. In five revolutions, signals would have reached a distance
roughly 3 times $r_0$ outwards. Furthermore, the radial flows near
both boundaries are likely to be subsonic, thus the boundary
conditions could in principle affect the flow inside.
We have tried several different choices with
$r_1=0.2, 0.4$, $r_2=2.0, 3.2$, in various combinations.
The dynamic behavior of the flow near $r_0$ 
(say $[0.5-1.5] r_0$) is mostly independent of the size of the disk, 
as shown partly by the linear theory (Paper II) and the 
following simulation results. Using a large $r_2/r_0$ ratio
is relatively simple since the dynamics at large $r$ is 
smooth and evolves slowly. Having a large $r_0/r_1$ ratio,
however, is obviously difficult. The strong Keplerian shear
will continuously shorten the radial wavelength of radial
propagating perturbations so that it becomes impossible to 
resolve them at late time.  In most runs presented here,
we have used $r_1/r_0 = 0.4$ and $r_2/r_0 = 2.0$, though 
we have also made many runs with $r_1/r_0 = 0.2$ and 
$r_2/r_0 = 3.2$ to ensure that similar results are obtained.

After many tries, we have determined two types of radial boundary
conditions that give reasonable results. One type of boundary
condition is simply setting all the quantities in the ghost
cells to be same as their initial values.
This roughly mimics the condition that the two 
fluxes on the ghost
cell boundaries are the same so that the mean quantities at the cell
centers are not changed. Note that this condition still allows
the material to flow off the grid since the flux is calculated
at the cell interfaces. 
Another type of boundary condition for obtaining the ghost cell
quantities is to fix
$v_\phi$  and to extrapolate the {\em variations} between successive
timesteps for the other three variables. The idea is to mimic the
passage of ``weak signals'' assuming that the variations are
typically small. We find that both types of boundary conditions
work quite well on the outer boundary and there is negligible
reflection from the outer boundary. At the inner boundary, however,
we do not believe we have found a proper (or the most ideal) 
boundary condition, if there is one. The strong shear (i.e.,
short radial wavelength), the subsonic motion, and the fact that
variations are incident with an angle into the inner boundary
make it very difficult to eliminate reflections completely.
Consequently, the mass flux through the boundary might not be
quantitatively accurate, even though we have always observed mass
flowing out of the grids in all our runs.  

Another important issue is the role of
numerical dissipation. As discussed previously and further in the Appendix,
we have employed a hybrid scheme that uses a first-order method
(thus more dissipative) in regions with shocks and sharp discontinuities. 
The question is whether the transport we observed from the simulations
is dominated by numerical dissipation or true physical effects.
It is difficult to directly measure the amount of numerical
dissipation. We have performed several tests. First, for a given
initial equilibrium without the perturbations, we are able to evolve
the disk to a time of $> 50$ revolutions at $r_0$ with the disk
staying in the same initial equilibrium. For
example, the maximum radial velocity (normalized by $v_\phi(r_0)$)
found is less than $10^{-6}$, compared to zero initially. 
Another set of tests we did was to look
for convergence of azimuthally averaged quantities at different times
using different numbers of grid points for the radial ($nr$) and 
azimuthal ($np$) directions, including
$nr=100, 200,$ and $400$, and $np = 200, 400,$ and $800$. 
We typically find that one can use the
$200\times 200$ grid for a quick run out to 20 orbits at $r_0$ 
($\sim 20$ minutes on a Dec Alpha machine) with reasonably good results,
and use the $400\times 400$ grid for high resolution runs ($\sim 2.5$ hrs 
out to 20 orbits). 
One further test was
 to compute the fraction of Lax-Friedrichs (LF) flux (first-order
accurate) versus Lax-Wendroff (LW) flux (second-order accurate). 
This fraction is
zero when shocks are not present (i.e., pure LW flux is used). 
For runs with strong shocks, this fraction can
reach up to a few percent briefly and locally in the disk. 

We want to emphasize that the {\em physics} of the instability
we will discuss later does not depend on the boundary condition, at
least not critically. This is very different from some other global
hydro instabilities, such as the Papaloizou \& Pringle
instability. The different treatments of the boundaries, however, do
result in some minor quantitative differences, e.g., in estimating the 
transport efficiencies. Similarly, these estimates are affected by the
numerical dissipation as well. Nevertheless, we are confident in the
general physics
presented here, such as how and where the angular momentum is
transported and how the matter flows. But we are less confident in 
some of the exact numbers presented. Perhaps more sophisticated numerical
schemes and better boundary conditions can improve this situation.    

\subsection{Diagnostics}

The timescale of the simulations is referenced by the orbital period at
$r_0$ (i.e., time $t = 2\pi$ stands for one revolution at $r_0$). 
For most runs, we are able to run the simulations out to time $t =
126$ which is 20 orbits at $r_0$. This translates into $\sim 80$ orbits at
$r_1 = 0.4$ and $\sim 7$ orbits at $r_2 = 2.0$. This gives the
system ample time for the nonlinear interactions to develop,
since the linear
growth stage usually takes $<8$ orbits (see below). Depending on the
amplitude of the maximum $\langle v_r \rangle$, this duration
also spans a few local accretion timescales,
allowing us to probe the accretion dynamics.
As we will discuss later, 20 orbits are probably enough since the
radiative cooling, which is not included here, is expected to play an
important role after this many revolutions.  
 
Most variables are normalized by their values at $r_0$, such as 
radius $r/r_0$, angular velocity $\Omega/\Omega_k(r_0)$, growth rate 
$\gamma/\Omega_k(r_0)$, and density 
$\Sigma/\Sigma(r_0)$. From these, the sound speed is normalized by the
Keplerian velocity 
$v_{\phi k}(r_0)$ and its value is typically $\sim 0.1$. The corresponding
pressure is then $\sim \Sigma c_s^2/\Gamma \sim 6\times 10^{-3}$ for
$\Gamma = 5/3$. So, if the velocity variations are close to $0.1$, shocks
are expected to occur.

Besides displaying sequences of global 2D distributions of various physical
quantities, the time-evolution of the disk can also be studied through
various azimuthally averaged variables. One key quantity is the radial
angular momentum transport due to the $r-\phi$ component of the
Reynolds stress $\langle \Sigma \delta v_r \delta v_\phi \rangle$, 
which is usually set
equal to $\alpha P$ in the Shakura-Sunyaev formulation, where $\alpha$
is a dimensionless parameter characterizing the angular momentum
transport efficiency. We will discuss this further in the
presentation of our results.

\section{Results: Comparison with the Linear Theory}
\label{sec:result1}

\subsection{Confirmation of the linear theory}

The dependence of RWI on various initial equilibrium disks with
different ${\cal A}$ and azimuthal mode
number $m$ has been given in linear theory in Paper II.
In this subsection, we show that the linear theory is confirmed very
nicely by our nonlinear simulations. 
For a given initial equilibrium with a specific ${\cal A}$, we can
use the eigenfunction of a specific unstable mode (i.e., $\delta \Sigma,
\delta v_r, \delta v_\phi, \delta P$ from the
linear theory) as the initial small amplitude perturbations. 
The linear theory predicts that the
unstable mode should grow exponentially with a certain growth rate
$\gamma$ and mode frequency $\omega_r$. Furthermore, as  shown in
Paper II, the unstable modes are global, so that an exponential growth
should occur throughout the whole disk. This behavior will certainly
change when the nonlinear effects become dominant.

The upper panel of Figure \ref{fig:lt} shows the early evolution of  
radial velocities at
3 different radii, $r/r_0 = 0.7, 1$, and $1.3$. 
The initial small perturbations are based on an 
eigenfunction with $m=5$ and ${\cal A}=1.4$ for an NGB equilibrium. 
The linear theory gives $\omega_r = 4.95$ and $\gamma = 0.243$. 
It is clear that the unstable mode is
exponentially growing throughout the whole disk with the same growth
rate and mode frequency. A rough estimate from this figure gives  
$\omega_r \approx 4.92$ and $\gamma \approx 0.24$,
which are essentially the same values as
predicted by linear theory.
In the lower panel of Figure \ref{fig:lt} we present a run using 
an NGB equilibrium with ${\cal A}=2.5$ and $m=5$. The predicted 
$\gamma$ is $0.61$ and $\omega_r$ is $4.9$. The estimated values from
the nonlinear simulation are again in perfect agreement.

The nearly perfect confirmation of the linear theory also
serves as a good test of our nonlinear code as it resolves 
the mode and captures its exponential growth. Similar results
are obtained for other types of equilibria as well as different
azimuthal mode number $m$, which we do not present here.

\subsection{Local Axisymmetric Instability vs. RWI}

One key result from the linear theory studies (Paper II) is that there
exists a range of ${\cal A}$ where the disk is unstable to RWI but
{\em stable} to the local axisymmetric instability. For ${\cal A}$ larger
than certain critical value (which depends on the details of disk
initial equilibrium), in addition to RWI, the disk has a small region
where $\kappa^2 + N^2$ is less than 0,
where $\kappa$ and $N$ are the epicyclic frequency and the radial
Brunt-V$\ddot{a}$is$\ddot{a}$l$\ddot{a}$ frequency 
due to the radial entropy variation, respectively. This makes the disk
also susceptible to local axisymmetric instabilities according to the  
Solberg-Hoiland criterion (Paper II). In actual disks, the
disk evolution will depend on which instability has a higher growth
rate. We suspect that RWI will likely play an important role
regardless. This is because the growth rate of RWI is usually already quite
high ($> 0.3 \Omega(r_0)$) for large ${\cal A}$. Even if an axisymmetric
instability grows first, since the quantity $\kappa^2 + N^2$ is less
than 0 only in a very small region near $r_0$ (see Figure 5 in Paper
II), the instability acts to stabilize this region,
but might still leave a finite ${\cal A}$ from which
RWI can grow. Furthermore, since RWI is a global mode (compared to the
axisymmetric instability which is local), its impact on the disk
dynamics could be much larger. In some of our simulations that have
a localized $\kappa^2 + N^2 < 0$, it seems that RWI is always the dominant
instability and in fact we have never detected any deviation
from the exponential growth of RWI. We conclude that local
axisymmetric instability is not important in our studies.

\section{Results: Nonlinear Stage}
\label{sec:result2}

Even though we are only dealing with 2D disk simulations, the
nonlinear evolution of the flow is quite complicated. We will show
in this section that large scale structures, such as vortices and 
in some cases, shocks, are produced.
Besides the initial exponential growth due to the linear instability,
even more localized pressure variations are produced both in the radial and
azimuthal directions which feed back to the original instability. 
So, the disk evolution enters into a somewhat self-feeding state
during which significant transport of mass and angular momentum 
is observed.

\subsection{Different Types of Runs}

\begin{table}
\caption{The initial setup of all 13 runs, which are categorized
by homentropic or nonhomentropic Gaussian bump (HGB and NGB) and
homentropic or nonhomentropic step jump (HSJ and NSJ). Different
bump/jump amplitudes (${\cal A}$) and widths ($\Delta r/r_0$) are
represented.  
All the runs except one (T13) have used eigenfunctions from the 
linear theory (Paper II) as the initial small amplitude perturbations, 
which are quantified by the azimuthal mode number
$m$ and growth rate $\gamma$ (normalized by $\Omega(r_0)$).
Run T13 uses random initial perturbations.
\label{tab:runs}}
\begin{tabular}{|c|c|c|c|c|c|}
\tableline
runs & Type & ${\cal A}$ & $\Delta r/r_0$ & $m$ & $\gamma$ \\
\tableline
T1 & HGB & 1.12 & 0.05 & 3  & 0.10 \\
T2 & NGB &  1.22 & 0.05 & 5 & 0.11 \\
T3 & HSJ & 0.4 & 0.05 & 5  & 0.11 \\
T4 & NSJ & 0.3 & 0.05 & 5 & 0.11 \\
\tableline
T5 & HGB & 1.35 & 0.05 & 3 & 0.27 \\
T6 & NGB & 1.55 & 0.05 & 5 & 0.32 \\
T7 & HSJ & 1.2  & 0.05 & 5 & 0.29 \\
T8 & NSJ & 0.8  & 0.05 & 5 & 0.30 \\
\tableline
T9 & HGB & 1.6 & 0.1 & 3  & 0.10 \\
T10& HGB & 2.4 & 0.1 & 3  & 0.32 \\
\tableline
T11 & HGB & 1.17 & 0.05 & 3 & 0.15 \\
T12 & HGB & 1.25 & 0.05 & 3 & 0.20 \\
\tableline
T13 & NGB & 1.51 & 0.05 & & \\
\tableline
\end{tabular} 
\end{table}

We have performed a large number of runs using various initial
equilibria. We choose to present 13 runs. Their
properties are summarized in Table \ref{tab:runs}. 
All 13 runs use $nr\times np = 400\times 400$. 
The first 12 runs have $0.4 \leq r/r_0 \leq 2.0$ and 
are perturbed (from equilibrium) 
using their respective eigenfunctions with 
a specific azimuthal mode number $m$. 
All these simulations
are run to time $t=126$, i.e., 20 orbits at $r=r_0$.
The last run, T13, has
$0.2 \leq  r/r_0 \leq 2.0$ with random initial perturbations.
It is run  to time $t=200$ (i.e., $\sim 32$ turns).

Runs T1-T4 represent 4 types of initial equilibrium with roughly
the same linear growth rates ($\gamma \sim 0.1$) and similarly 
for runs T5-T8 ($\gamma \sim 0.3$). Runs T1-T4 all
have small ${\cal A}$ so that $\kappa^2 + N^2$ 
is everywhere positive (i.e., only RWI is present).
Runs T5-T8, however, have a narrow region with $\kappa^2 + N^2 < 0$,
though RWI seems to be the only instability present.
Figure \ref{fig:allpre_low} shows the evolution of pressure
for runs T1-T4 and Figure \ref{fig:allpre_high} is for runs T5-T8.
In order to make the pressure variations more clearly visible, 
we have actually plotted $r^{3/2}P$ for runs of NGB and NSJ types
to take away the $r^{-3/2}$ dependence in the background pressure
(the second and fourth rows in Figures \ref{fig:allpre_low} and
\ref{fig:allpre_high}). 
The left column shows the initial pressure distribution,
the middle column is at the time when the linear instability
just saturates ($t=3,$ and $7$ orbits for lower and higher growth rate
runs, respectively), and the last column is at 
time of $t=20$ orbits. 

The initially axisymmetric pressure distribution
has broken up and became nonaxisymmetric with distinct, 
organized regions, which turn out to be vortices. 
In addition, large scale spiral arms around these vortices
have developed. It is clear that these nonaxisymmetric 
features, such as the hot spots in pressure, are quite persistent.
Furthermore, runs T5-T8 are evolving at a much higher rate than
that of runs T1-T4, especially in the nonlinear regime as well. 
This can be seen by comparing the middle and right columns in
Figures \ref{fig:allpre_low} and \ref{fig:allpre_high}.
Note that the pressure in the inner region ($r/r_0 < 1$) shows 
an overall increase as shown in Figure
\ref{fig:allpre_high}. This will be discussed in detail in
later sections as we believe that this is a clear signature of
overall accretion.



Figures \ref{fig:allpre_low} and \ref{fig:allpre_high} 
also show that 
the dynamic behavior of the disk can be roughly divided into three
regions: near $r_0$ (say, $r/r_0 \sim 0.8-1.2$), inner ($r/r_0 < 0.8$) 
and outer ($r/r_0 > 1.2$) parts. Vortices and shocks form near $r_0$, 
and they constantly generate waves that propagate towards both the 
inner and outer parts of the disk. (We have confirmed that these
waves propagate at the sound speed.) These sound waves, being
continuously sheared by the background flow, develop into spiral waves
that might eventually lead to shocks. 

In all runs, the disk evolution can be roughly
divided into three stages: an exponential growth of small 
amplitude perturbations, the formation of 
vortices, around which shocks are sometimes produced,
and the global mass and angular momentum transport.  
The exponential
growth phase has been discussed previously. We now
discuss the rest of evolution in detail.

\subsection{Formation of vortices}
\label{sec:vortform}

In this subsection we take a closer look at the formation of
vortices. We specifically study two runs, T1 and T5, since
they have relatively simple initial configurations, such
as constant entropy. 
Figure \ref{fig:vort_global} shows a global view of the whole
disk with vortices. The pressure distribution is color-coded and
the overlaying arrows map out the flow patterns
around $r_0$ after subtracting $v_{\phi}(r_0, t=0)$ 
(i.e., in the comoving frame
that has a azimuthal velocity of $v_{\phi~K}(r_0, t=0)$). 
The upper panel is taken from T1 at a time of 7 orbits and
the lower panel is from T5 at 3 orbits. 
Both runs are initialized with the $m=3$ unstable mode.
Even though the Keplerian
shear is still the dominant flow pattern,
vortices are clearly formed in the flow. 


One dominant feature of these vortices is that the vortical motion
is anticyclonic (the ``spin'' axis of the vortex is opposite to the disk
rotation axis) and the vortex
encloses a localized high pressure region.
The nonuniform pressure distribution along the 
azimuthal direction (i.e., the $-\partial P/\partial \phi$ term) 
is the main driving force in the formation of vortices. 
Such nonaxisymmetry grows out of RWI directly (see Figure 1
of Paper II). 
Taking the flow at $r/r_0=1$ from T5 as an example, in Figure
\ref{fig:pvrvp}, we have plotted the evolution of pressure $P$
(upper panel), azimuthal velocity $v_{\phi}$ (middle panel)
and radial velocity $v_r$ (lower panel) for times $t=0, 1, 2, 3$
orbits, which are represented by the solid, dotted, dashed,
and long-dashed lines, respectively. 
The flow is initially
rotating with a single $v_\phi$ that balances the gravity plus
pressure gradient.  As the
pressure gradient $-\partial P/\partial \phi$ builds up due to RWI,
the azimuthal velocity $v_\phi$ will decrease (increase) if the 
$-\partial P/\partial \phi$ force is negative (positive). A decreasing
(increasing) $v_\phi$ causes the flow to move inwards (outwards) 
radially since gravity and rotation are no longer in balance. 
So, surrounding each localized high pressure region, an
``anticyclone'' is formed.
In fact, this is probably the only vortical flow pattern that could
survive in this nearly Keplerian shear flow.

The development of 
nonlinearity is clearly seen from these curves as well.
At early times the azimuthal variations are still sinusoidal
but become strongly concentrated by the time of 3 orbits 
(most clearly seen in the long-dashed curve for pressure, 
for example). Interestingly, even though the flow is nonlinear,
these vortices remain azimuthally separated and are moving with the
same speed around the disk (cf. Figures \ref{fig:allpre_low} and
\ref{fig:allpre_high} at 20 orbits).
Later we will discuss situations when vortex merge does occur.


\subsection{Vortex Radial Width and Shocks}

Figure \ref{fig:svortex} shows a closeup view of one vortex 
from those in Figure \ref{fig:vort_global}, but now 
in the $\{r, \phi\}$ plane instead. The upper row is for
T1 and the lower for T5 with pressure $P$ on the left
and entropy change $\Delta S = \ln(P/\Sigma^{\Gamma})-S_0$ 
on the right. Note that for both T1 and T5, the whole disk is set up 
with a single entropy $S_0 = \ln(P_0/\Sigma_0^{\Gamma})$.



As the flow is still predominantly Keplerian, there is
a fundamental constraining effect on the radial width of a 
vortex.  The Keplerian 
shear flow implies that the relative azimuthal flow speed 
($v_{\phi}(r_2) - v_{\phi}(r_1)$)
exceeds the local sound speed when 
\begin{equation}
\label{eq:width}
|r_2-r_1|/r_0 \geq 2 c_s/v_\phi(r_0), 
\end{equation}
which is roughly $\sim 0.2$ in our case. In other words,
imagine having a rod with a total length of $0.2 r_0$ and placing
it radially with its mid-point at $r_0$, the two ends of the rod
could still communicate at the sound speed. This will not be true,
however, for structures with larger radial width. 
In fact, this perhaps is the basis for the long-held belief that
it is extremely difficult to maintain a long-lived vortex in accretion
disks since it will be sheared away.

The radial width of the vortices produced in runs T1 and T5 can
be best estimated from Figure \ref{fig:strm_t1t5}, where
the streamlines are depicted for the same
flow regions as in Figure \ref{fig:svortex}.
The vortex from T1 indeed stays within the limit imposed by
equation (\ref{eq:width}) but the vortex from T5 is rather large.
By ``large'' we
mean that the vortex has a ``core'' region that has a radial extent 
of nearly $0.4 r_0$. It also has extended ``arms'' (high pressure 
regions, cf. Figure \ref{fig:svortex}) that go out much farther 
in radial extent. The flow around this vortex is obviously much 
more complicated than that around the vortex from T1. 

The critical difference lies in the fact that shocks are produced
around the vortex in T5 but not in T1. The vortex in
T5 is surrounded by 
4 ``arms'' which are labeled as A, B, C, and D in 
Figure \ref{fig:strm_t1t5} (see the corresponding locations in
Figure \ref{fig:svortex}). These arms
mark the places where the pressure is high, so are the
density and temperature (not shown here).
Arms A and C are clearly shocks that have a pressure jump of nearly
a factor of 2 and strong entropy production (lower right plot
of Figure \ref{fig:svortex}). Note that arms A and C start at 
the radial location $r \approx 0.9, 1.1 r_0$, respectively,
agreeing precisely with that expected from equation (\ref{eq:width}). 
Arms B and D are, however, not shocks. Instead, they are the
location where the rarefaction wave from shocks at arms A and C
meets with the background flow. This is supported by the fact that
no entropy variation is seen even though the pressure variation is
close to a factor of $1.5$. In addition, the streamlines are smooth
near arms B and D (unlike A and D), the apparent flow direction change
at arms B and D are actually due to an imbalance of
gravity, rotation and pressure forces, not from a shock.
The flow is expanding very strongly in the region between arms B and C, 
as well as between A and D, where the flow directions are strongly 
altered and the radial velocity reaches its
maximum for both the infalling and outward motion (indicated by
the length of the arrows in Figure \ref{fig:svortex}).
We believe that these flow structures, especially the shocks, 
are pivotal in the
formation and ``protection'' of the vortices against the background
shear flow. They have enclosed a region within which the flow is 
subsonic and the streamlines are closed.  

In contrast, no shocks are produced around the vortex in T1,
as evidenced by the extremely small change in entropy
shown in the upper right plot of Figure \ref{fig:svortex}.
(Note the different scales of $\Delta S$ in the two entropy
plots.)  These
changes at $\sim 10^{-4}$ level are most likely due to the numerics
alone. In other words, the whole disk remains homentropic to a
high degree. 

\subsection{Dependence on the initial width $\Delta r/r_0$}

We have also investigated the vortex radial size dependence on 
the initial pressure bump width $\Delta r/r_0$. This is done
in runs T9 and T10, where $\Delta r/r_0 = 0.1$. These two
runs are designed to have the same linear growth rates as
runs T1 and T5, respectively. A very similar evolution is observed
in these two runs, and Figure \ref{fig:sv_wide} shows the
flow velocities around a single vortex together with its pressure 
distributions in the $\{r, \phi \}$ plane for runs T9 (left)
and T10 (right). They are taken at times of $7, 4$ orbits,
respectively.  The lower panels are shown at a time of 20 orbits.

The vortex in T9 has a small radial width ($< 0.2 r_0)$), the
same as that in T1. Consequently, no shocks are observed either.
For the vortex in T10,  shocks are again observed at very
similar locations as those in T5. This can be seen by
comparing the upper right plot of Figure \ref{fig:sv_wide}
with the lower left plot of Figure \ref{fig:svortex}, the two
shock structures (arms A and C) are quite similar, with the same
starting radial locations at $\sim 0.9$ and $1.1 r_0$, respectively.
The compressions due to rarefaction waves, however, are not
as pronounced in T10 as those in T5. 
Thus the radial width of a vortex does not depend
on the initial pressure bump width. The characteristics of the 
Keplerian flow is the dominant factor in determining the vortex 
width, with or without shocks.


\subsection{Radial Drift of Vortices}

The right panel of Figure \ref{fig:sv_wide} also reveals an 
interesting phenomenon: there is a slight but noticeable 
inward radial drift of the vortices between the times shown 
(4 and 20 orbits).  At 20 orbits, this drift is only visible in 
run T10 but not in T9, presumably because T10 is evolving much
faster. The amount of radial drift appears to be small, but 
it actually implies a high accretion rate. Using the usual scaling
relation for the radial accretion velocity, $v_r \sim \alpha c_s (H/r)$,
where $H$ is the disk vertical scale-height, we get
\begin{equation}
\alpha \sim \frac{v_r}{v_\phi}~\frac{v_\phi}{c_s}~\frac{r}{H}
\sim \frac{\Delta r_{\rm drift}}{2\pi r_0 N}~(\frac{v_\phi}{c_s})^2~.
\end{equation}  
\noindent where $N$ is the number of orbits at $r_0$, and
$v_\phi/c_s \approx 10$. Reading from the right column of
Figure \ref{fig:sv_wide}), we get
$\Delta r_{\rm drift} \approx 0.05 r_0$ and $N = 20$, implying
that $\alpha \sim 0.04$. This simple estimate turns out to 
be quite consistent with more detailed analysis presented later.

\subsection{Dependence on the initial perturbations}

In physical systems, the initial perturbations are unlikely 
to be a single eigenfunction given by the linear theory, though 
one can obviously decompose the variations into various eigenmodes.
In run T13, we use a random small amplitude initial perturbation
(this ensures nonaxisymmetry by default).
An initial exponential growth is again observed (not shown here).
In Figure \ref{fig:merge}, we show 12 snapshots of the disk in 
color-coded radial
velocity, with the first 11 frames at $t=0,2,4,6,...,20$, and the
last frame at $t=32$ orbits. 
Vortices are clearly produced, just as in all the other runs we
have presented. There is a clear trend that vortices merge with
each other, going from $\sim 5 - 6$ vortices initially to only one 
dominating vortex at $\sim 16$ orbits. The fact that the $m=5, 6$ 
modes grow first  is expected from the linear theory 
analysis as they have the highest linear growth rates 
(cf. Figure 10 of Paper II).  Note that even though the vortices
are nearly corotating with the background flow, there is nevertheless
a difference in the phase velocity for different $m$ modes.
Eventually a faster moving vortex will catch up with a slower 
one and the two vortices will merge. In the end, there is only one vortex
left in the system since, given enough time, 
any slight difference in the phase velocity
will lead to an interaction between two vortices. 
We emphasize that such strong interactions are due to the
fact that vortices are excited/produced at nearly the same radius and 
the radial drift of these vortices are very slow. In
a real system where multiple ``bumps'' might be present at different
radii, multiple vortices could be present.    


\subsection{Mass and angular momentum transport}

We now address the critical question of 
the mass and angular momentum transport in these disks.
Neglecting dissipation (e.g., shocks) for a moment,  
angular momentum is conserved (except for the loss due to the flow 
through the boundaries), but
may be redistributed as the disk evolves. 
Following the treatment in Balbus \& Hawley (1998), we can separate
the velocities into a mean component and a ``varying''
component, $v_\phi(t) = \overline{v}_{\phi}(t) + \delta v_\phi$
and $v_r(t) = 0 + \delta v_r(t)$, where $\overline{v}_{\phi}(t)$ is
obtained by averaging $v_\phi(r,\phi,t)$ over $\phi$ at 
a particular radius $r$ and time $t$.  Consequently,
the radial flux of angular momentum is decomposed into two
parts
\begin{equation}
\label{eq:fang}
r^2 [\overline{v}_{\phi} \langle \Sigma v_r \rangle + 
\langle \Sigma v_r \delta v_\phi \rangle ]~,
\end{equation}
where $\langle ... \rangle$ indicates averaging over $\phi$,
$\int d\phi/2\pi$. The first term indicates
the direct radial flow of matter and is proportional to the
mass accretion rate, $2\pi R \langle \Sigma v_r \rangle$.
The second term represents the radial angular momentum transport 
through the {\em correlations} of velocity component variations.
Traditionally this has been thought of as the turbulent Reynolds
stress $\langle T_{r \phi} \rangle \equiv \langle \Sigma v_r 
\delta v_\phi \rangle$, whose origin has been the subject of 
intensive research for decades. Furthermore, as emphasized in 
Balbus \& Hawley (1998), what is more important is the 
{\em positive correlation} between
$v_r$ and $\delta v_\phi$ instead of the mere presence or amplitude 
of these variations. Even though we do not regard the 
flow we are studying as
turbulent, the same requirement, i.e., the positive correlation
between $v_r$ and $\delta v_\phi$, still holds the key to an outward
transport of angular momentum. 

The most important result of the vortices generated by RWI is
that the flow pattern around these vortices is
perhaps an ideal configuration for an outward angular momentum
transport process. 
This is due to the fact that the azimuthal pressure gradient 
causes variations in $v_\phi$ and consequently leads to the 
generation of $v_r$ via radial force balance, i.e.,
a decrease (increase) in $v_\phi$ leading to a negative (positive) 
$v_r$ (see detailed discussion in \S \ref{sec:vortform}).
Such a correlation directly ensures the radial angular momentum flux
via transport (cf. equation (\ref{eq:fang})) is positive, i.e.,
an outward transport of angular momentum.

To quantify the crucial role of vortices in angular momentum
transport, we define a 2D version of a modified 
$\alpha$ coefficient,
\begin{equation}
\label{eq:alp1}
\alpha_{ij} = \frac{\Sigma_{ij}~ v_{r ij}~ [ v_{\phi ij} - 
\overline{v}_\phi(r,t) ]}{P_{ij}}~,
\end{equation}
where the indices $\{ij\}$ stand for $\{r\phi\}$. Here, we use
$\overline{v}_\phi(r,t)$ as the ``mean'' background flow (though it
is a debatable choice), but it nevertheless
gives a good indication as to which regions/structures are
contributing most importantly to the angular momentum transport.
Figure \ref{fig:svort_alp_t5} shows
the strength and distribution 
of $\alpha_{r\phi}$ around a vortex in the $\{r,\phi\}$ plane
for run T5 at time of 3 (left panel) and 20 orbits (right panel). 
Similarly, Figure \ref{fig:svort_alp_t1} shows the 
distribution of $\alpha_{r\phi}$ at 7 (left panel) and 20
orbits (right panel), respectively, for run T1.

Comparing Figures \ref{fig:svortex} and \ref{fig:svort_alp_t5}, 
one can see that strong outward angular momentum transport occurs
in the expansion regions behind shocks A and C, with localized
$\alpha_{r\phi}$ exceeding $0.1$. Similar structures are observed in
results from T1  even though the amplitude is at a much
reduced level and shocks are not present.
Even after 20 orbits,  the main features and their strength 
(notice the scaling) remain amazingly steady and strong. 
These vortices are rather remarkable in this regard. 


To further quantify the global transport efficiency, we 
can take an azimuthal average of equation (\ref{eq:alp1}), which
is equivalent to 
\begin{equation}
\label{eq:alp2}
\langle\alpha \rangle = \langle T_{r\phi}/P \rangle~,
\end{equation}
where $\langle T_{r \phi} \rangle = \langle \Sigma \delta v_r 
\delta v_\phi \rangle$ and 
\begin{equation}
\langle \Sigma \delta v_r \delta v_\phi \rangle = \langle \Sigma v_r
v_\phi \rangle - \langle \Sigma v_r \rangle \langle \ell \rangle/r~~,
\end{equation}
\noindent where $\langle \ell \rangle = \langle \Sigma\ell \rangle/
\langle \Sigma \rangle$ is the average specific angular
momentum (Hawley 2000).
The lower panels of Figures \ref{fig:svort_alp_t5} and
\ref{fig:svort_alp_t1} give $\langle \alpha \rangle$ at the same 
times as their corresponding upper panels. 
The transport efficiency of T5 is much higher than that of T1,
by a factor of at least 30 (though there is only a factor of 3
difference in linear growth rates). So, it is not surprising that 
the evolution of T5 is much faster than that of T1, as observed
previously.

To illustrate the dynamics of transport efficiency,
we present the evolution of $\langle \alpha \rangle$ for runs T1-T8
in Figure \ref{fig:alp_all}. The strength of transport usually reaches
a peak when the vortices first form (the dotted lines), but
settles down to maintain a steady level, and so there is
relatively little difference between 10 and 20 orbits.
Even though the transport of angular momentum in the disk has 
both outward and inward components (as indicated by the positive and
negative values of $\alpha_{r\phi}$ in Figures \ref{fig:svort_alp_t5} 
and \ref{fig:svort_alp_t1}), on average
the angular momentum is transported outward through each radial 
``ring'',  as indicated by the predominantly positive
$\alpha$ given in Figure \ref{fig:alp_all} for all the runs.
This is extremely encouraging and perhaps
the most important result of this study.


Another important point is that the transport can be roughly divided 
into two different regions, that associated with vortices and that
associated with the trailing spiral waves that are present in both the 
inner and outer parts of the disk. The physics behind the outward angular 
momentum transport by trailing spiral waves is actually quite
similar to what we have discussed previously for the transport 
by vortices, since the azimuthal pressure gradient is 
fundamentally responsible
for causing the positively correlated velocity variation components.
In some sense, a vortex is just a much more pronounced nonlinear
manifestation of such transport processes. 
In the case that the vortex is radially large and strong (i.e., run
T5), the transport strength around the vortex is much larger 
($\alpha \sim 10^{-2}$) than that of the spiral wave region
($\alpha \sim 10^{-3}$). In the case that the vortex is weak
(i.e., run T1), they become comparable ($\alpha \sim 3\times
10^{-4}$). The increase of $\alpha$ at smaller radii is related
to the stronger shear, though it is difficult to accurately estimate 
it. 

Note that the spiral waves are produced by continuously
shearing the radially propagating sound waves that are 
generated in the vortex region. This is one of the important 
features of the linear theory (Paper II) in which waves are allowed 
to ``tunnel through'' the trapping region where the vortex is 
produced. The consequence of this connection is that 
transport will occur not only near the position of the initial 
bump/jump, but also throughout the disk as a whole, thus giving rise
to a much larger, global impact. 

We have also investigated other physical quantities
using azimuthal averages. These quantities include the pressure
$\langle P \rangle$, mass surface density $\langle \Sigma \rangle$, 
mass flux (accretion rate) $F_\rho = \int d\phi r\Sigma v_r$, and
angular momentum flux $F_j = \int d\phi r\Sigma r v_\phi$.
They are shown in Figures \ref{fig:lplot_t1} and 
\ref{fig:lplot_t6t8}, for runs T1, T6 and T8,
respectively. 

There are several generic features that appear in all 
these runs:

(1) As expected, the instability always tries to remove the 
bump/jump. The lower linear growth rate runs evolve more
slowly than those
with higher linear growth rate runs. The nonlinear
saturation levels are also different, as evidenced by the magnitude
of mass accretion rate $F_{\rho}$, for example, in Figures
\ref{fig:lplot_t1} and \ref{fig:lplot_t6t8}.

(2) As a clear confirmation of the efficient accretion that is
going on in the disk flow, the pressure in most parts of the disk 
is increasing with a variable amount 
(except the initial bump/jump, of course). This is especially
true for the inner part of the disk.
We believe that this increase is from the release of 
gravitational energy due to the ``global'' accretion caused
by both vortices and spiral waves. This effect seems inevitable
since we did not include any cooling effect in our equation of
state.

(3) The global nonlinear evolution brings an additional
lack of axisymmetry and radial variations in the disk flow.
This is manifested in the average pressure distributions
where large radial as well as azimuthal (cf. Figures
\ref{fig:allpre_low} and \ref{fig:allpre_high})
gradients are produced. These strong gradients
will be susceptible to the same Rossby wave instability
we are studying. It is then not very surprising that the
system can sustain itself for a long time, consistent
with our results. 



\section{Discussion}
\label{sec:discuss}

One advantage of our nonlinear simulations is that we are 
guided by a robust linear instability that has been investigated 
previously. The precise confirmation of the linear theory
not only validates the presence of this instability, but also 
provides a check for our nonlinear codes. Consequently, 
some of the usual concerns 
with numerics are not as important. 
At the fully nonlinear stage when shocks are present, it is,
however, difficult to capture all the dissipation perfectly.
So we are less confident in some of the exact numbers presented,
but we believe that the large scale structures of this instability
have been captured correctly. Furthermore, by following the
instability evolution through the linear growth stage, we 
have gained more confidence on the physical mechanism of the
instability and have singled out the key physical processes in the
nonlinear regime, such as the formation of vortices and shocks. 
There are still a number of physical issues that deserve further
discussion.

\subsection{Setting Up the Initial Equilibrium}

In realistic astrophysical situations, the initial conditions will certainly
be system dependent. The idealized bump/jump along with the background
disk described by the present studies 
might arise in the close binary systems where matter
tends to be stored at the large radii first; Or it could be the radiation
heating from the central star that causes a localized region of
the disk to be hotter; Or it could be the edge between a proto-planet
and its surrounding disk material. 
By investigating the role of RWI
in many different initial configurations, we show that
RWI is very robust. Quite generally, the disk is potentially 
unstable due to RWI whenever there are ``bumps, 
edges, clumps'' present in the disk.

\subsection{Dependence on the amplitude ${\cal A}$ and Physics
of Saturation}

Since there is a clear difference between runs with low and high
growth rates, we now compare 4 runs,
T1, T11, T12 and T5, all of which use a Gaussian bump but with an 
increasing amplitude ${\cal A}$. 
Their linear growth rates range from $0.1$ to $0.27$. Instead of
showing global distributions of various quantities, we opt for 
a single quantity, the maximum radial velocity 
$|v_r(r_0,\phi,t)|$ at $r_0$, to describe the development
of the instability. This is plotted in Figure \ref{fig:allvr}
for different runs as a function of time. Along with these 4 runs,
we also plot the results from runs T6, T7, and T8.
While it seems that all runs with high growth rates eventually
saturate at roughly the same ${\rm max}(|v_r|)$, the saturation
level of ${\rm max}(|v_r|)$ for lower growth rate runs clearly
depends on the linear growth rate. 


Since the value of ${\rm max}(|v_r|)$ can be regarded as a
rough measure of the level of nonlinearity  
and the associated transport 
efficiency, one question that naturally arises is ``what is the
physics causing the instability growth to saturate?'' 
There are several possibilities. One is that when the linear growth
is slow, saturation is achieved by the removal
of the bump/jump, i.e., the driving of the instability. 
We believe this is what happens when the bump/jump is small,
such as runs T1-T4. The evolution shown in Figure \ref{fig:lplot_t1} 
supports this interpretation (see also Figure \ref{fig:allpre_low}). 
As shown before, anti-cyclonic vortices are formed
surrounding regions with high  pressure and density, 
but there are no shocks in the flow and the flow entropy is well 
conserved (cf. Figure \ref{fig:svortex}). 

A relevant study on this issue can be found in Laughlin et
al. (1997, 1998) with a somewhat different setting. From their
excellent nonlinear mode coupling analyses in self-gravitating 
gaseous disks, they concluded that the growth of the dominant unstable
mode can modify the background disk profile so as to prevent its
further growth, causing saturation without relying on dissipation.

When the bump/jump gradually increases, the instability growth becomes
fast enough that the saturation is achieved both by the
removal of the bump/jump and the formation of shocks. 
Anti-cyclonic vortices are again formed with their radial width 
being roughly 4 times the thickness of the disk. The larger radial
size leads to the formation of shocks, which in turn limit
the growth of radial velocities, causing the instability to saturate.

\subsection{Energy Conversion and Dissipation}

One of the key physical links in accretion disk physics is to
understand how gravitational energy is converted into internal
energy, part  of which can then be radiated away. The 
increase of internal energy, however, can occur either 
adiabatically (i.e., entropy is conserved) or with 
dissipation (i.e., heat generation with increasing entropy). 
In the classical, geometrically thin and optically thick $\alpha$ 
disk model (similar to the physical condition we are considering 
here), the disk is assumed to be axisymmetric, Keplerian and 
quasi steady. The angular momentum transport is via
an assumed anomalous viscous stress that is related to the Reynolds
stress $\langle \Sigma v_r \delta v_{\phi} \rangle$.
These conditions lead to a relation where, at
large radii, local heat production (via dissipation) can be a 
factor of 3 times the available gravitational energy release, 
a direct result of viscous transport (cf. the textbook by Frank, King 
\& Raine 1985). The viscous heating rate per unit volume per
gram is $dQ \sim \nu (rd\Omega/dr)^2$ (Lynden-Bell \& Pringle 1974).
Estimating $\nu \approx \alpha c_s H$, where $H$ is the thickness
of the disk, we would expect an entropy increase
$dS$ by relating $TdS = dQ$,
\begin{equation}
\label{eq:ds}
dS \approx (9/4)~(\Omega t)~\alpha~~,
\end{equation}
i.e., the entropy increases linearly with time 
as transport continues.

Our results in run T1, however,
seem to indicate a different route in the energy conversion process.
The evolution in T1 satisfies the
adiabatic condition to a high degree,
as evidenced by the near constancy of entropy of the whole disk,
which maintains its initial value (cf. Figure \ref{fig:svortex}). 
(In fact,
we have written a different version of the code by requiring that
the entropy of the flow be conserved, as contrasted with the code we
presented which uses the total energy conservation. Both codes
give very similar results.) This disk, however, is transporting
angular momentum outward with an equivalent $\alpha$ of 
$10^{-4}$ (conservatively), as shown by Figure \ref{fig:svort_alp_t1}.
Using equation (\ref{eq:ds}), this level of angular momentum
transport would imply an entropy increase far greater
than what we have obtained, which is less than  
$dS \approx 2\times 10^{-4}$ as shown in upper right plot of
Figure \ref{fig:svortex}. This special case proves an important 
physical point, that angular momentum can be transported outward
in a disk {\em without} dissipation. The released gravitational 
energy goes entirely to
$PdV$ work, which is done adiabatically.
Again the studies by Laughlin et
al. (1997, 1998) are relevant here. In their case, 
the nonlinear mode interactions create a nonsteady perturbing 
potential that continuously drives the disk evolution, giving
angular momentum and mass transport without dissipation. 
  
Whether the real astrophysical disk operates via   
a ``maximal heat/entropy production'' route or an adiabatic route
is unclear, and observational constraints have been scarce. 
Different energy conversion processes, however, do predict different 
amounts of energy that
can be radiated and consequently give different radiation spectra.
It is needless to say that the global disk structure and evolution
obtained from our study differ fundamentally from the classical
$\alpha$ disk model, especially in how energy
is converted and how heat is generated. For example, one obvious
difference is that the transport efficiency has a radial dependence,
which implies that the disk is always in a dynamic state and 
can be more properly described as ``surges'' of
matter accretion. How to meaningfully construct global accretion
disk models under these conditions and  relate them
to observations might be very interesting.

\subsection{Late Time Evolution and Effects of Radiative Cooling}

We are able to run most simulations out to $40-50$ revolutions 
(at $r_0$) and find that disks are continually evolving, though we
have only presented the results up to 20 orbits. 
In fact, in some cases the disk has evolved so much 
that we believe that we should
not run those simulations much longer than 20 revolutions.
This is because additional physical effects that are not included
in the present study could become too important to ignore. One such 
effect is the radiative cooling. For example, as shown in Figure
\ref{fig:lplot_t6t8}, 
the average inner disk pressure has gone up by a large factor 
(i.e., $\times 2$), while the density shows only relatively small variations. 
This implies a large change in disk 
temperature, which could mean a 
large change in radiative loss as well.

As emphasized in Paper II, the disk has to be relatively hot 
(i.e., $c_s/v_\phi \geq 0.05$) in order to have a ``healthy'' growth 
rate for RWI. The increase
in disk pressure helps to sustain the dynamic evolution. On the
other hand, ignoring radiation means that the cooling time of disk
should be relatively long compared to the disk rotation periods. This
requirement, as discussed in Paper II, implies a minimum column
density of the disk matter so that heat can be trapped inside the 
disk for several revolutions. 
Since our simulations are 2D, we
could not directly model the effects of radiative cooling with respect
to the vertical transport. One way to circumvent this difficulty is to
add an ad hoc local cooling function that removes the internal energy
at a specified rate (see R\'o\.zyczka \& Spruit 1993). Similarly, 
we have tried
to add a loss term in the energy
equation as $-e/\tau_c$, where $e = P/(\Gamma - 1)$ is the internal
energy and $\tau_c$ is the characteristic cooling timescale.
The parameter $\tau_c$ is likely to be a complicated function of
density, temperature (hence radius of the disk) and radiative
transport processes. As a simple approximation, we have tried to
relate $\tau_c$ to the local Keplerian rotation period by a single
constant. Indeed we find the trend that the shorter the cooling time,
the weaker the nonlinear effects of the instability.
Intuitively, if the cooling time is shorter
than one rotation period, then changes caused by the vortex motion
(i.e., compressions and expansions occurring during a ``thermal''
cycle) will likely be damped out very quickly, thus strongly limiting
the efficiency of any transport processes. 

\subsection{2D versus 3D}

Another important aspect is the 3D nature of the disk
flow. The 2D approximation is expected to break down in several
ways. The effects of strong shear are clearly visible in all the
runs (cf. the inner region of Figures \ref{fig:allpre_low} and
\ref{fig:allpre_high})  where the spiral waves
become wound tighter and tighter with increasing orbital velocity.
Such a short radial wavelength situation probably violates the 2D
assumption. The radial propagation of the sound waves itself can be
weakened by the ``refraction'' effect discussed in Lin et al. (1990),
thus limiting its impact radially. 
Furthermore, as the disk expands/contracts both vertically and
radially as pressure varies, dissipation by the irreversible 
processes becomes inevitable (e.g., expansion into a near vacuum). 
This will probably prove to be
the most important 3D effect, though detailed 3D simulations
are needed to address this problem quantitatively.

\subsection{Entropy gradient is not necessary}

It is worthwhile emphasizing that RWI grows under a wide variety of
physical conditions and that it is {\em not} necessary to have an entropy  
variation in the disk, at least for 2D disks (Paper II). This is
supported by the above results (run T1, for example). 
What drives the mode unstable is the steep pressure gradient which
gives rise to the ``trapping'' physics that allows the mode to
be amplified. The amplification of the unstable mode is the result of
repeatedly passing through the corotation radius that is 
residing in a ``trapping'' region (Paper II).

Entropy variations in the disk could, however, introduce more features
since the additional potential vorticity can be driven
thermodynamically by
the $\nabla P \times \nabla \Sigma$ term. In actual disks, entropy
variation might be inevitable as argued by Klahr
\& Bodenheimer (2000). 
Further studies are needed in order to better quantify the
effects of entropy variations.

\section{Conclusions}
\label{sec:conclu}

We have studied the global nonlinear evolution of the Rossby wave
instability in a nearly Keplerian flow, following our
previous linear theory analysis (Lovelace et al. 1999; Li et al. 2000).
During the linear growth stage of the instability, our nonlinear 
simulations agree extremely well with the linear theory results 
(e.g., the growth rate and mode frequency).  

In the nonlinear stage, we have shown that vortices naturally form, 
enclosing a high pressure and density region. These vortices are 
(nearly) corotating with the background flow but are 
counter-spinning (i.e., anti-cyclones). We have elucidated 
the physical mechanism for the production of these vortices, 
namely through the azimuthal pressure gradients, 
and shown that they are long-lived structures within disks.
These vortices are shown to be extremely important for
transporting angular momentum outward and for causing global
accretion. In fact, by analyzing the flows around each vortex,
we have shown why they are the ideal ``units'' for outward
angular momentum transport, namely by giving rise to 
``positively'' correlated radial and azimuthal velocity variations,
i.e., $\langle \Sigma v_r \delta v_{\phi} \rangle > 0$.  
Shocks are formed when the instability is strong and these
shocks limit the radial extent of a vortex to be less than 4
times the thickness of the disk. 
Furthermore, trailing spiral waves are produced both interior
and exterior to the vortex region. 
These trailing spiral waves are produced by shearing the radially
propagating sound waves generated by vortices, and they
serve as an additional means of transporting angular momentum
outwards. 

We find that the angular momentum transport efficiency is not
a constant throughout the whole disk; it has both radial and azimuthal
dependences, and evolves continuously. Consequently, 
we recognize the need to construct
global models of accretion disks that reflect such 
nonaxisymmetry and dynamics. 

Several important physical issues are, however, not addressed
in this paper. Three-dimensional modeling is especially needed 
to address 
the issue of radiative cooling and how much heat/dissipation 
is produced locally. In addition, combining this hydrodynamic 
instability with those associated with magnetic fields will
be very interesting as well.    

\acknowledgements{
We are indebted to N. Balmforth, P. Goldreich, J. Finn, E. Liang, 
and R. Lovelace
for many useful discussions. H.L. gratefully acknowledges the 
hospitality of UC-Santa Cruz during the final completion of the paper
and would like to thank D. Lin and L. Margolin for carefully reading the
manuscript and providing many thoughtful comments.
We thank the referee for pointing to us the papers by Laughlin et al.
H.L. also acknowledges the support of an Oppenheimer Fellowship.
B.B.W. has been supported by CHAMMP program at LANL.
R.L. has been supported in part by the Czech Grant Agency grant
201/00/0586 and would like to thank the Institute for Geophysics and
Planetary Physics (IGPP)  for hosting his visit at LANL.
This research is supported by the Department of Energy,  
under contract W-7405-ENG-36.
}

\appendix

\section{Numerical details}

Our numerical method differs from that used by Masset (2000) in several
ways:

(1) We do a Galilean transformation of the split angular
equations, transforming to a coordinate system rotating with
a constant velocity at each radius. The velocity is chosen to
be as close as possible to the mean angular velocity such that
the transformation back to the fixed system involves only a
shift of indices. 

(2) The stability condition is the standard 
Courant, Friedrichs, and Lewy (CFL) condition 
computed from the radial velocity and the sound speed, with
limit 1. The time interval determined in this way is still too
large for the angular integration, in spite of the reduction in
angular velocity obtained from the co-moving system, but we use
partial time-stepping in angle, satisfying the angular CFL
condition at each partial step. 

(3) All the sources are included in the radial sweep; the sources
are not done in a split step.

(4) A simple hybrid scheme is used for both the angular and radial
integrations. The idea is to use a weighted combination of the
second-order Richtmyer two-step version of the Lax-Wendroff (LW) scheme
and the two-step first-order Lax-Friedrichs (LF) scheme, with weights chosen
to favor LF in regions possibly containing shocks, and to favor
LW in smooth regions, while maintaining the second-order accuracy
there. This is an idea first proposed in \cite{hz72}. There are many
variations of this method that are described very well with references
to the original literature in \cite{l98}. The particular weights we
have chosen seem to provide a robust scheme.

In dimensional splitting we consider  separately the radial equations
with the sources;
\bea
\label{eq:radial}
\pdt w 
+ \pdr f(r,w)  
+ S(r,w) = 0,
\eea
and the angular equations
\bea
\label{eq:angular}
\pdt w 
+ \pdp g(r,w) = 0.
\eea
A single cycle consists of first a determination of the time interval
$\dt$,
then a sequence spanning two time steps of
the form radial-angular-angular-radial. More formally, we would have
calls to routines of the form
\begin{center}
CALL radial(w,w')
\end{center}
where the input w is the vector of dependent variables at time $t^n$,
and w' is the output. Then using w' as input,
\begin{center}
CALL angular(w',w)
\end{center}
where w is now the result of one time step.

\subsection{The radial integration}
Equal radial intervals $\dr$ define equally spaced radii $r_i,i=
0,N+1$, presumed at the center of radial cells. The cell vertices
are at $r_{i+\hlf}$. The inner radial boundary is at $r_{\hlf}$,
the outer at $r_{N+\hlf}$.

In describing the radial integration we suppress the dependence on
angle: the difference equations being described must also be applied
at each angle. The equations are written in flux form
\bea
w'_i=w_i-\lambda(F_{i+\hlf}-F_{i-\hlf})-\dt S^*,
\eea
\bea
\lambda=\dt/\dr,
\eea
with the flux $F$ and the source $S^*$ defined below.

The flux is a hybrid:
\bea
F=\alpha F^{LF}+(1-\alpha) F^{LW},
\eea
\beann
0\leq \alpha \leq 1.
\eeann
$F^{LF}$ is a Lax-Friedrichs flux:
\bea
F^{LF}_{i+\hlf}=-\frac{1}{2\lambda}(w_{i+1}-w_{i})
+\frt (f(w_{i+1})+f(w_i))+\hlf f(w^*_{i+\hlf}),
\eea
\bea
\label{eq:rhalf}
w^*_{i+\hlf}=\hlf (w_{i+1}+w_i)-\frac{\lambda}{2}
(f(w_{i+1})-f(w_i))-\frac{\dt}{2}{\hat S}_{i+\hlf}.
\eea
$F^{LW}$ is the Lax-Wendroff flux
\bea
F^{LW}_{i+\hlf}=f(w^*_{i+\hlf}).
\eea
The source $S^*$ is a vector with four components, $S^{m*},
m=1,4$, but only two are nonzero, $S^{1*}=S^{3*}=0$: 
\bea
S^{4*}_i=\hlf (S^{4*}_{i+\hlf}+S^{4*}_{i-\hlf})
\eea
with
\bea
S^{4*}_{i+\hlf}=(\Sigma v_r)^*_{i+\hlf}/r_{i+\hlf},
\eea
and,
\bea
S^{2*}_{i}&=&-\hlf [(\Sigma ((v_{\phi})^2-1/r)^*_{i+\hlf}
+(\Sigma ((v_{\phi})^2-1/r)^*_{i-\hlf}]\nn\\
&&+r_{i+\hlf}(P(w^*_{i+\hlf})-P(w^*_{i-\hlf}))/dr.
\eea
Turning to the predictor step equation(\ref{eq:rhalf}), we found 
it necessary
to make an adjustment. The greatest stumbling block to
a long run, say ten turns of the disk, is the appearance of a
negative pressure somewhere in the disk. The final set of difference
equations we have used does not have this defect {\em without setting
any artificial lower bound on the pressure}, at least in all the
many runs we have done. The problem is that apparently minor
changes in the scheme can cause negative pressures. Thus, we found
that using the angular momentum conserving form in the predictor
step did not work. Of course, angular momentum is still conserved,
so there should be no objection on that ground.

The weight $\alpha$ is
\bea
\alpha=\left[\min (\frac{|(v_r)_{i+1}-(v_r)_{i-1}|}{c_i},1)\right]^2~~,
\label{eq:weight}
\eea
where $c_i$ is the local sound speed.

\subsection{The angular integration}
The angular integration finite-difference equations have the same
hybrid form as those for the radial integration, except that
there are no sources, and $v_{\phi}$ replaces $v_r$
in (\ref{eq:weight}).
Since the radius is constant on any one angular
sweep, all extraneous radial factors can be canceled. The hybrid
weights are computed using angular velocity differences rather than
radial velocity differences. However, as indicated in section
\ref{sec:num}, a local co-moving frame and partial time-stepping are
used.

Equal angular intervals $\dphi$ define equally spaced angles $\phi_j,j=
1,M$, presumed at the center of angular cells. The cell vertices
are at $\phi_{j+\hlf}$, with  $\phi_{\hlf}=0$, 
$\phi_{M+\hlf}=2\pi$. The periodicity is enforced by means of the
indexing, that is, the index $j$ is always understood
as $j+M-{\rm int}((j-1+M)/M)M$.

In describing this method, we suppress the radial dependence. The
local co-moving frame is determined by first defining the average
angular velocity, following Masset.
\bea
{\bar v_{\phi}}=\frac{1}{M}\Sigma_1^M (v_{\phi})_j.
\eea
The actual velocity shift used is given by
\bea
\tau=\frac{{\bar v_{\phi}}}{r}\frac{\dt}{\dphi}
\eea
\bea
l={\rm int}(\tau+\hlf)
\eea
\bea
v_0=lr\frac{\dphi}{\dt}
\eea

The purpose of this shift is to reduce the effect of the large angular
velocity on the time step for the angular integration. If the angular
velocity were independent of $j$, the use of ${\bar v_{\phi}}$ would
accomplish this at the expense of having to do an interpolation
later, since the shift of the coordinate system in one time step
would not be an integer number of cells. By using $v_0$ it
will be an integer shift
Then, at the beginning we define new conserved variables by replacing
$v_{\phi}$ with $v'_{\phi}=v_{\phi}-v_0$. This entails a replacement of
$\Sigma v_{\phi}$ by $\Sigma v'_{\phi}$ and of $\Sigma E$ by 
$\frac{P}{\Gamma-1}+\hlf \Sigma ( (v'_{\phi})^2 + v_r^2)$.
At the end of the angular sweep we will have computed new variables,
$\Sigma_j,(v_r)_j,(v_{\phi})_j$, and $P_j$, but these are in the
moving coordinate system. To get back to the fixed system, first
replace $(v_{\phi})_j$ by $(v_{\phi})_j+v_0$, then shift the
density, velocities and pressure from cell $j$ to cell $j+l$,
then redefine the conserved variables.

For the actual integration we cover one time interval $\dt$ with an
integer number of steps with interval
\bea
\dt'=\frac{\dt}{n}
\eea
where $n$ is determined from the angular stability condition. Thus,
if
\bea
\Sigma=\max_j(\frac{|v'_{\phi}+c|}{r},\frac{|v'_{\phi}-c|}{r})
\eea
then
\bea
n={\rm int}(\frac{\Sigma \dt}{\dphi}) +1
 \eea

We have found that the average $n$ is about 1.2.


\begin{figure}
\caption{The exponential growth of the radial velocity during
the linear growth stage of the Rossby wave instability. 
Two runs, both of NGB type, are shown with different 
bump amplitude ${\cal A}$. The dotted curves show the magnitude
of radial velocities at three fixed locations in the observer's frame
at $r = 1, 0.7, 1.3$ and $\phi = 0$, from top to bottom, respectively.  
For a given ${\cal A}$ and $m$ ($=5$ for both panels), 
the linear theory of RWI predicts a specific
mode frequency $\omega_r$ and growth rate $\gamma$. All the
curves are from our simulations and from each curve, one can
get $\omega_r$ from the oscillations (the dotted line)
and $\gamma$ from the slope of its ``envelop'' (the solid line). 
Both quantities show excellent agreement with
the predictions of the linear theory.  
\label{fig:lt}}
\end{figure}

\begin{figure}
\caption{The evolution of the pressure for runs T1-T4 (lower
growth rate runs). 
Each row consists of snapshots of the whole
disk at three 
different times of each run (T1-T4 from top to bottom). 
From left to right, $t=0, 7, 20$ orbits, 
respectively. The color code is for pressure, which is in units
of $10^{-3}$. Note that the scale is different
for each run. Each run is initialized using small amplitude perturbations
based on the eigenfunction of its linear instability with a
specific azimuthal mode number $m$, which is $m=3,5,5,5$ from
top to bottom. The pressure of T2 and T4 (second and fourth rows)
has been multiplied by $r^{3/2}$ in order to make the pressure 
variations more easily visible. Isolated hot spots (high pressure)
are clearly visible, and they are the centers of large anticyclonic
vortices. Large scale spiral arms are produced in connection 
with these vortices as well.
\label{fig:allpre_low}}
\end{figure}

\begin{figure}
\caption{Similar to Figure \ref{fig:allpre_low} except using
results from runs T5-T8 (high growth rate runs). The middle column 
is now
at a time $t=3$ orbits and the right column is again at 20 orbits.
The amplitude of the pressure variations is larger than that
seen in Figure \ref{fig:allpre_low}. The pressure has clearly
increased in the inner part of the disk. Vortices and large scale
spirals are produced as well.
\label{fig:allpre_high}}
\end{figure}

\begin{figure}
\caption{Vortices in a disk. Pressure is color-coded (in units
of $10^{-3}$). Arrows indicate the flow pattern near $r_0$ in
a comoving frame of $v_{\phi}(r_0)$. Vortices are anticyclonic,
enclosing high pressure regions. Large-scale spirals are produced
as well, in connection with the vortices. The upper panel is
a snapshot from T1 at $t=7$ orbits and the lower panel is
from T5 at $t=3$ orbits. Both runs use an $m=3$ unstable mode,
which is why there are three (nearly) corotating vortices.  
\label{fig:vort_global}}
\end{figure}

\begin{figure}
\caption{The production process of vortices. Shown is how the pressure
(upper panel), the azimuthal velocity (middle panel) and the radial
velocity (lower panel) vary along the azimuthal ($\phi$) direction 
as the instability develops. The results are from run T5 and are taken
at $r = r_0$. The solid, dotted, dashed and long-dashed curves 
in each panel are at $t = 0, 1, 2, 3$ orbits, respectively. 
Note the $\pi/6$ shift between the peaks in pressure and peaks in 
$v_\phi$. Such correlation is derived from the fact that 
the azimuthal pressure gradient $-dP/d\phi$ is responsible for
the variations in $v_\phi$. Consequently, the imbalance
between gravity, rotation and radial pressure gradient in the radial
direction introduces radial motion. This explains why
the largest positive (negative) radial velocities occur when
$v_\phi$ is the largest (smallest), i.e., super- (sub-) Keplerian.  
\label{fig:pvrvp}}
\end{figure}

\begin{figure}
\caption{A closeup view of one vortex in the $\{r,\phi\}$ plane. 
The upper two plots show one vortex from run T1 at $t=7$ orbits.
The flow near $r_0$ is indicated 
by the arrows, together with the color-coded pressure (left plot) and 
the color-coded entropy change $\Delta S$ (right plot).
Similarly, the lower two plots show one vortex from run T5 at
$t=3$ orbits with the pressure (left) and entropy change (right)
distributions.  The vortex from
T1 is relatively weak with small radial motions.
No shocks are present, as shown by the exceedingly small variations 
in entropy $\Delta S$.
The vortex from T5, on the other hand, is very strong with significant
radial motion. There are 4 pressure ``arms'' (locations
with high pressure) surrounding this vortex, as indicated by
the labels A, B, C, and D. Shocks are formed at arms A and C, as shown
by the large increase in entropy. Arms B and D are probably
not shocks, instead, they are produced by the rarefaction waves from the 
shocks at A and C. Consequently, entropy does not change at B and D.
The high pressure band and the high entropy band in the lower left
region of the two T5 plots (lower panel) are from the arm A of
another vortex (not shown). 
\label{fig:svortex}}
\end{figure}

\begin{figure}
\caption{The streamlines around the same vortices from T1 (left) and
T5 (right) as those shown in Figure \ref{fig:svortex}.
The vortex from T1 has a small radial width ($< 0.2 r_0$) 
but the vortex from T5 has a large radial width ($\sim 0.4 r_0$).
The sharp changes in streamline direction at arms A and C further prove
that they are indeed shocks, whereas arms B and D are not.
\label{fig:strm_t1t5}}
\end{figure}

\begin{figure}
\caption{A closeup view of vortices from run T9 and T10, both
of which have a wider initial bump width $\Delta r/r_0 = 0.1$. 
The pressure is color-coded in all plots and the comoving flow
velocities are overlayed as arrows.
The two left panels are from T9 at $t=7$ (upper) and $20$ (lower)
orbits, respectively. The two right panels are from T10 at 
$t=3$ (upper) and $20$ (lower) orbits, respectively.
Again, shocks are formed around the vortex from T10, limiting
its radial width to be less than $\sim 0.4 r_0$. 
There is a noticeable inward radial drift of the vortex from T10.
A high angular momentum transport efficiency is implied
from such a drift.
\label{fig:sv_wide}}
\end{figure}

\begin{figure}
\caption{Merge of vortices. Shown are 12 snapshots from run T13, where
random initial perturbations also lead to large-scale vortex
formation. From top left to bottom right, the frames are from
time $t=0, 2, 4, 6, ..., 20$ and $t=32$ orbits, respectively. 
The radial velocity is color-coded and the same color scale is 
used for all frames. The amplitude of the radial velocity grows
from small values ($\sim 0$) to nearly sound speed ($\sim 0.1$).
Each pair of blue (in-fall) and red (out-moving) regions 
indicates one vortex. Initially there are 5-6 vortices present
in the disk since these modes have the highest linear growth rates
(Paper II). But these vortices have slightly different phase
speeds going around the disk, which means that vortices will
eventually catch up with each other and interact. In the end, 
only one strong vortex left. (A lower resolution version
is shown here in order to reduce the file size.)      
\label{fig:merge}}
\end{figure}

\begin{figure}
\caption{Angular momentum transport by vortices. Shown are the 
spatial distribution of $\alpha_{r \phi}$ 
(equation [\ref{eq:alp1}])
in the $\{r,\phi\}$
plane around a vortex from run T5 at time $t=3$ (left) and $20$
(right) orbits. The parameter $\alpha_{r \phi}$ is color-coded.
The lower panel shows the azimuthally averaged $\langle 
\alpha_{r,\phi} \rangle$ (equation [\ref{eq:alp2}])
as a function of radius, at the
same times as those in the upper panel. Positive 
$\langle \alpha \rangle$ indicates an outward transport 
of angular momentum. The transport is peaked at the vortex
region and remains finite away from the vortex.
The strength of the transport still
remains high at $20$ orbits. 
\label{fig:svort_alp_t5}}
\end{figure}

\begin{figure}
\caption{Similar to Figure \ref{fig:svort_alp_t5} except
using run T1 at $t=7$ (left) and $20$ (right) orbits.
The overall strength of $\langle \alpha \rangle$
is much smaller than that from T5.  
\label{fig:svort_alp_t1}}
\end{figure}

\begin{figure}
\caption{The radial dependence and evolution of 
$\langle \alpha \rangle$ (equation [\ref{eq:alp2}])
for runs T1-T8 (top left to 
bottom right). The solid, dotted, dashed and long-dashed
curves are at time $t=0, 7, 10, 20$ orbits for runs T1-T4
and $t=0, 3, 10, 20$ orbits for runs T5-T8, respectively. 
Each row uses the scale shown on the left.
The angular momentum, on average, is always transported
outwards through each radius (ring). The transport peaks
when the vortices first form but remains steady
between 10 and 20 orbits.
\label{fig:alp_all}}
\end{figure}

\begin{figure}
\caption{The evolution of the azimuthally averaged physical
quantities for run T1. From top to bottom, 
these are pressure $\langle P \rangle$, 
surface density $\langle \Sigma \rangle$, radial mass flux
(accretion rate) $F_{\Sigma}$, radial angular momentum flux
$F_j$, and angular momentum transport efficiency 
$\langle \alpha \rangle$. The solid, dotted and dashed
curves in each plot are at time $t=0, 10, 20$ orbits, respectively.
The disk evolution is relatively slow, with small changes
in averaged quantities (but cf. Figure \ref{fig:allpre_low}
for changes in azimuthal direction). 
\label{fig:lplot_t1}}
\end{figure}

\begin{figure}
\caption{Similar to Figure \ref{fig:lplot_t1} but for runs T6 
(left) and T8 (right). The solid, dotted, dashed and long-dashed
curves in each plot are at time $t=0, 3, 10, 20$ orbits, respectively.
The disk evolution is much faster and stronger compared to run T1. 
\label{fig:lplot_t6t8}}
\end{figure}

\begin{figure}
\caption{The growth and saturation of radial velocities for
various runs. Runs T1, T11, T12 and T5 have the same initial
configurations except for the increasing amplitude  ${\cal A}$.
The saturation level increases as the linear growth rate increases,
but ceases to do so when the linear growth rate is large enough,
as shown by runs T5, T6, T7 and T8 (note the difference in their
peak values).
This is because shocks are responsible for the growth saturation
in all the high growth rate runs.
\label{fig:allvr}}
\end{figure}

\end{document}